\documentclass[prd,showpacs,amsmath,amssymb]{revtex4}
\usepackage{hyperref,slashed,graphicx,color}
\begin{document}
\title{The Effects on $S$, $T$, and $U$ from Higher-Dimensional Fermion Representations}
\author{Hong-Hao Zhang}
\affiliation{%
Department of Physics, Tsinghua University, Beijing 100084, China}

\author{Yue Cao}
\affiliation{%
Department of Physics, Tsinghua University, Beijing 100084, China}

\author{Qing Wang}
\affiliation{%
Department of Physics, Tsinghua University, Beijing 100084, China}

\begin{abstract}
Inspired by a new class of walking technicolor models recently
proposed using higher-dimensional technifermions, we consider the
oblique corrections from heavy non-degenerate fermions with two
classes of higher-dimensional representations of the electroweak
gauge group itself. One is chiral SM-like, and the other is
vector-like. In both cases, we obtain explicit expressions for $S$,
$T$, $U$ in terms of the fermion masses. We find that to keep the
$T$ parameter ultraviolet-finite there must be a stringent
constraint on the mass non-degeneracy of a heavy fermion multiplet.
\end{abstract}
 \maketitle

\section{Introduction}
Despite its tremendous success, the standard model (SM) has several
drawbacks. On the one hand, the Higgs particle has not yet been
found in experiments; on the other hand, the SM suffers the
hierarchy problem and triviality from a theoretical point of view.
Thus, the SM may not be correct, or at least it is just an effective
theory at the electroweak scale. There are many new physics
possibilities beyond the SM. Although we do not know whether nature
really behaves like one of them or not, we can estimate their
effects on the current electroweak precision measurements. Peskin
and Takeuchi's $S$, $T$, $U$-formalism is a practical way to do this
job \cite{Peskin:1990zt}. Since the current SM parameter fits
indicate that $S$ and $T$ are small negative numbers, and $U$ is
also close to zero \cite{Yao:2006px}, those new physics models which
give large positive contributions to $S$ and $T$ are presumably
excluded. Thus, the oblique correction parameters $S$, $T$, $U$ are
often used to judge whether a new model is compatible with
experiments or not. If the SM is not a full theory, there will be
new heavy particles above the electroweak scale. Provided the new
particles feel the electroweak interactions, they should give
corrections to $S$, $T$, $U$ whether they are fermions, scalars, or
gauge bosons.

Recently, there has been increasing interest in a new class of
walking technicolor models, using technifermions with
higher-dimensional, rather than fundamental, representations of the
technicolor gauge group \cite{Sannino:2004qp}. Their walking
dynamics feature can avoid unacceptably large flavor changing
neutral currents. If these models were true, in general it will be
also possible for the presence of heavy fermions with
higher-dimensional representations of  the electroweak gauge group
$SU(2)_L\times U(1)_Y$ itself. Of course, these particles could give
corrections to $S$, $T$, and $U$. In an earlier paper by Dugan and
Randall \cite{Dugan:1991ck}, the effects to the $S$ parameter from
general fermion representions of $SU(2)_L\times SU(2)_R\times
U(1)_Y$ has been considered assuming a strict custodial $SU(2)_C$
symmetry. Later, the corrections to $S$, $T$, $U$ and also to
triple-gauge-vertices from a heavy non-degenerate fermion doublet
has been estimated respectively \cite{Peskin:1990zt, Inami:1995ep}.

In this paper, we will calculate the corrections to $S$, $T$, and
$U$ from two classes of higher-dimensional fermion representations
of $SU(2)_L\times U(1)_Y$ itself. One is the SM-like chiral type, in
which right-handed fermions are singlets, while left-handed fermions
form a multiplet of the $SU(2)_L\times U(1)_Y$ group. The other is
the vector-like case, in which the left and right-handed fermion
multiplets transform the same way under the electroweak group. In
the following, the strict custodial symmetry will be relaxed to an
approximate symmetry so as just to keep the $T$ parameter
ultraviolet-finite. In each case, we obtain a mass constraint on a
fermion multiplet to satisfy this demand. At the end of the paper a
brief concluding remark is given.

\section{The SM-like Chiral Representations}
Consider a SM-like heavy fermion multiplet with $N=2j+1$ dimensions
and with quantum numbers of $SU(2)_L\times U(1)_Y$ as
\begin{eqnarray}
&&\psi_L=\begin{pmatrix}\psi_{j}\\ \psi_{j-1} \\ \vdots\\ \psi_{-j}\end{pmatrix}_L
\sim (2j+1, Y)\,,\nonumber\\
&&\psi_{l,R}\sim (1, Y+l)\,,\qquad (l=j,\,\ldots,\,-j)\,.
\end{eqnarray}
For simplicity, we restrict this electroweak multiplet to be a color
and technicolor singlet. If it is also a color multiplet, the effect
on the result is just a multiplying factor of the number of colors.
In general, this $N$-plet is non-degenerate, and we denote their
masses $m_l$ where the subscript $l$ runs from $-j$ to $j$. They
couple to the electroweak gauge bosons via
\begin{eqnarray}
\frac{e}{\sqrt{2}s}(W_\mu^+J_+^\mu+W_\mu^-J_-^\mu)
+\frac{e}{cs}Z_\mu(J_3^\mu-s^2J_Q^\mu)+eA_\mu J_Q^\mu\,,\label{int}
\end{eqnarray}
where $c\equiv\cos\theta_W$, $s\equiv\sin\theta_W$, and
\begin{eqnarray}
&&J_+^\mu=\sum_{l=-j}^{j-1}\sqrt{(j-l)(j+l+1)}\bar{\psi}_{l+1,L}\gamma^{\mu}\psi_{l,L}\,,\nonumber\\
&&J_-^\mu=\sum_{l=-j+1}^{j}\sqrt{(j+l)(j-l+1)}\bar{\psi}_{l-1,L}\gamma^{\mu}\psi_{l,L}\,,\nonumber\\
&&J_3^\mu=\sum_{l=-j}^{j}l\;\bar{\psi}_{l,L}\gamma^{\mu}\psi_{l,L}\;,\qquad
J_Q^\mu=\sum_{l=-j}^{j}(l+Y)\bar{\psi}_{l}\gamma^{\mu}\psi_{l}\;.
\end{eqnarray}
By computing the vacuum polarization amplitudes for the $N$-plet
fermion, we obtain their contributions to the oblique correction
parameters $S$, $T$, and $U$. As expected, we find that $S$ and $U$
are always ultraviolet-finite, and they are
\begin{eqnarray}
S&\equiv&16\pi[\Pi^{\prime}_{33}(0)-\Pi^{\prime}_{3Q}(0)]
=\frac{1}{3\pi}\sum_{l=-j}^{j}\bigg[l^2-2\,l\,Y\,\log\big(\frac{m_l^2}{\mu^2}\big)\bigg]\;,\\
U&\equiv&16\pi[\Pi^{\prime}_{11}(0)-\Pi^{\prime}_{33}(0)]
=\frac{4}{\pi}\bigg[\sum_{l=-j}^{j-1}\frac{(j-l)(j+l+1)}{2}f_1(m_{l+1}^2,m_l^2)
-\sum_{l=-j}^{j}\frac{l^{2}}{6}\log\big(\frac{m_l^2}{\mu^2}\big)\bigg]\;,
\end{eqnarray}
where $\mu$ is a mass scale parameter and the function $f_1$ is
defined as
\begin{eqnarray}
f_1(m_{l+1}^2,m_l^2)\equiv\int_0^1dx\,x(1-x)\log\bigg[\frac{xm_{l+1}^2
+(1-x)m_l^2}{\mu^2}\bigg]\;.\label{f1}
\end{eqnarray}
But the $T$ parameter can be generally ultraviolet-divergent, and
the result is
\begin{eqnarray}
T&\equiv&\frac{4\pi}{s^{2}c^{2}m_{Z}^{2}}[\Pi_{11}(0)-\Pi_{33}(0)]\nonumber\\
&=&\frac{1}{4\pi
s^{2}c^{2}m_{Z}^{2}}\bigg[A\big[\frac{1}{\epsilon}-\gamma-\frac{1}{2}+\log(4\pi)\big]
+\sum_{l=-j}^{j}2\,l^{2}m_{l}^{2}\log\big(\frac{m_l^2}{\mu^2}\big)
-\sum_{l=-j}^{j-1}(j-l)(j+l+1)f_2(m_{l+1}^2,m_l^2)\bigg]\;,
\end{eqnarray}
where the divergent term has been dimensionally regularized by
setting $\epsilon=(4-d)/2$, and the coefficient $A$ and the function
$f_2$ are respectively
\begin{eqnarray}
&&A\equiv\sum_{l=-j}^{j}(j^2+j-3\,l^2)m_{l}^2\;,\\
&&f_2(m_{l+1}^2,m_l^2)\equiv\int_0^1dx[xm_{l+1}^2+(1-x)m_{l}^2]
\log\bigg[\frac{xm_{l+1}^2+(1-x)m_{l}^2}{\mu^2}\bigg]\;.\label{f2}
\end{eqnarray}
In order to avoid the unacceptable disastrous divergence, the
coefficient $A$ must be zero. This relation gives a strong
constraint on the mass non-degeneracy of fermion multiplets. Since a
small value of the $T$ parameter is related to the approximate
custodial $SU(2)_C$ symmetry; if $A\neq 0$, the custodial symmetry
will be disastrously violated. Thus we call $A=0$ the custodial
symmetry soft-breaking condition. For $j=1/2$, ({\it i.e.} for a
fermion doublet,) this condition is satisfied automatically. For
$j=1$, the constraint is $m_{1}^2+m_{-1}^2=2m_0^2$. For a general
$j$, an interesting particular example respecting this constraint is
$m_l^2=m^2+l\Delta m^2$, for $l$ running from $-j$ to $j$.

\section{The Vector-like Representations}
Next, we consider a $(2j+1)$-dimensional vector-like fermion
multiplet as
\begin{eqnarray}
\psi_L=\begin{pmatrix}\psi_{j}\\ \psi_{j-1} \\ \vdots\\
\psi_{-j}\end{pmatrix}_L \sim (2j+1, Y)\;,\qquad
\psi_R=\begin{pmatrix}\psi_{j}\\ \psi_{j-1} \\ \vdots\\
\psi_{-j}\end{pmatrix}_R\sim (2j+1, Y)\;.
\end{eqnarray}
The interaction between these fermions and the electroweak gauge
bosons is of the same form as Eq.(\ref{int}), but now
\begin{eqnarray}
&&J_{+}^{\mu}=\sum_{l=-j}^{j-1}\sqrt{(j-l)(j+l+1)}\bar{\psi}_{l+1}\gamma^{\mu}\psi_{l}\,,\nonumber\\
&&J_{-}^{\mu}=\sum_{l=-j+1}^{j}\sqrt{(j+l)(j-l+1)}\bar{\psi}_{l-1}\gamma^{\mu}\psi_{l}\,,\nonumber\\
&&J_3^\mu=\sum_{l=-j}^{j}l\;\bar{\psi}_{l}\gamma^{\mu}\psi_{l}\;,\qquad
J_Q^\mu=\sum_{l=-j}^{j}(l+Y)\bar{\psi}_{l}\gamma^{\mu}\psi_{l}\;.
\end{eqnarray}
Likewise, we compute their contributions to $S$, $T$, and $U$
resulting in
\begin{eqnarray}
S&=&-\frac{2\,Y}{3\pi}\sum_{l=-j}^{j}l\,\log(\frac{m_l^2}{\mu^2})\;,\\
U&=&\frac{4}{\pi}\bigg[\sum_{l=-j}^{j-1}(j-l)(j+l+1)\bigg(f_1(m_{l+1}^2,m_l^2)
-\frac{1}{2}f_3(m_{l+1}^2,m_l^2)\bigg)
+\sum_{l=-j}^{j}\frac{l^2}{6}[1-2\log(\frac{m_l^2}{\mu^2})]\bigg]\;,\\
T&=&\frac{1}{4\pi
s^{2}c^{2}m_{Z}^{2}}\bigg[B\big[\frac{1}{\epsilon}-\gamma-\frac{1}{2}+\log(4\pi)\big]
-2\sum_{l=-j}^{j-1}(j-l)(j+l+1)f_2(m_{l+1}^2,m_l^2)\bigg]\;,
\end{eqnarray}
where the functions $f_1$ and $f_2$ have been defined in
Eqs.(\ref{f1}) and (\ref{f2}), and the function $f_3$ and the
coefficient $B$ are respectively
\begin{eqnarray}
&&f_3(m_{l+1}^2,m_l^2)\equiv
\int_{0}^{1}dx\frac{x(1-x)m_{l+1}m_{l}}{xm_{l+1}^2+(1-x)m_{l}^2}\;,\\
&&B\equiv\sum_{l=-j}^{j-1}(j-l)(j+l+1)(m_{l+1}-m_l)^2\;.
\end{eqnarray}
In this case, the custodial symmetry soft-breaking condition $B=0$
implies all the $m_l$'s must be equal, {\it i.e.}, this vector-like
multiplet must be degenerate, otherwise the custodial symmetry will
be unacceptably broken. But a mass-degenerate vector-like fermion
multiplet gives zero contribution to $S$, $T$, and $U$. Thus, any
custodial-symmetry preserved vector-like fermion representations
have no effect on the oblique correction parameters.

\section{A Concluding Remark}
In this paper, we have obtained the one-loop corrections to $S$,
$T$, and $U$ from two classes of higher-dimensional fermion
representations. When taking the fermion masses to be equal, our
expression for the $S$ parameter coincide with the PDG's result of
$S$ for degenerate fermions \cite{Yao:2006px}. When taking $j=1/2$
in the SM-like case, our expressions for $S$, $T$, and $U$ are
exactly those given in Ref. \cite{Peskin:1990zt}.

 We have shown that for the chiral case, in
order to keep $T$ ultraviolet-finite, there must be a constraint on
the mass non-degeneracy of the chiral multiplet. While for the
vector-like case, this constraint becomes even more stringent, and
it demands that vector-like multiplets must be degenerate, which
further implies that vector-like fermions cannot give any
contributions to $S$, $T$, $U$ as long as an approximate custodial
symmetry is imposed. These mass constraints may be potentially
useful for some model-building considerations.

Although the case of SM-like chiral representations we considered
above is just a special case where right-handed fermions are all
weak-singlets, it is sufficient to illustrate the point. A
generalization of this work to more general chiral representations
might be straightforward.

\begin{acknowledgments}
We would like to thank Prof. H.~J.~He and Dr. J.~K.~Parry for
helpful discussions. This work is supported by the National Natural
Science Foundation of China, and the Fundamental Research Foundation
of Tsinghua University.
\end{acknowledgments}


\begin{thebibliography}{}
\bibitem{Peskin:1990zt}
  M.~E.~Peskin and T.~Takeuchi,
  Phys.\ Rev.\ Lett.\  {\bf 65}, 964 (1990);

  M.~E.~Peskin and T.~Takeuchi,
  Phys.\ Rev.\ D {\bf 46}, 381 (1992).

\bibitem{Yao:2006px}
  W.~M.~Yao {\it et al.}  [Particle Data Group],
  J.\ Phys.\ G {\bf 33}, 1 (2006).

\bibitem{Sannino:2004qp}
  F.~Sannino and K.~Tuominen,
  Phys.\ Rev.\ D {\bf 71}, 051901 (2005);

  D.~K.~Hong, S.~D.~H.~Hsu and F.~Sannino,
  Phys.\ Lett.\ B {\bf 597}, 89 (2004);

  D.~D.~Dietrich, F.~Sannino and K.~Tuominen,
  Phys.\ Rev.\ D {\bf 72}, 055001 (2005);

  D.~D.~Dietrich, F.~Sannino and K.~Tuominen,
  Phys.\ Rev.\ D {\bf 73}, 037701 (2006);

  N.~D.~Christensen and R.~Shrock,
  Phys.\ Lett.\ B {\bf 632}, 92 (2006).

\bibitem{Dugan:1991ck}
  M.~J.~Dugan and L.~Randall,
  Phys.\ Lett.\ B {\bf 264}, 154 (1991).

\bibitem{Inami:1995ep}
  T.~Inami, C.~S.~Lim, B.~Takeuchi and M.~Tanabashi,
  Phys.\ Lett.\ B {\bf 381}, 458 (1996);

  M.~Tanabashi, Lecture given in {\it Frontiers of Particle Physics 2006: Beyond the Standard
  Model}, Beijing (2006).

\end{thebibliography}
\end{document}